\documentclass[journal,12pt,draftclsnofoot,onecolumn]{IEEEtran}

\usepackage{psfrag}
\usepackage{amsmath,amssymb}
\usepackage{mathrsfs}
\usepackage{graphicx,color,cite}

\newcommand{\Ts}{T_\mathrm{s}}
\newcommand{\Eq}{E_q}
\newcommand{\sinc}{\mathop{\mathrm{sinc}}\nolimits}

\newtheorem{theorem}{Theorem}
\newtheorem{lemma}[theorem]{Lemma}

\newtheorem{corollary}[theorem]{Corollary}
\ifCLASSINFOpdf
\else
\fi
\begin{document}
%
\title{Bandlimited Intensity Modulation}

 


%

%
\author{\IEEEauthorblockN{Mehrnaz~Tavan,
Erik~Agrell,
and~Johnny~Karout,~\IEEEmembership{Student Member,~IEEE}}\\
\thanks{
Research supported in part by the Swedish Foundation for Strategic Research (SSF) under
grant RE07-0026. The material in this paper was presented in part at the IEEE Global Communications Conference, Houston, TX, Dec.~2011.

M.~Tavan is with the Wireless Information Network Laboratory (WINLAB), Department
of Electrical and Computer Engineering, Rutgers University, Piscataway, NJ 08854 USA (e-mail: mt579@eden.rutgers.edu). E.~Agrell and J.~Karout are with the Department of Signals and Systems, Chalmers University of Technology, SE-412\,96 Gothenburg, Sweden (e-mail: agrell@chalmers.se, johnny.karout@chalmers.se).}}



%
%
%

\maketitle



%
\IEEEpeerreviewmaketitle

\begin{abstract}
In this paper, the design and analysis of a new bandwidth-efficient  signaling  method over the bandlimited intensity-modulated direct-detection (IM/DD) channel is presented. The channel can be modeled as a bandlimited channel with nonnegative input and additive white Gaussian noise (AWGN). Due to the nonnegativity constraint, standard methods for coherent bandlimited channels cannot be applied here. Previously established techniques for the IM/DD channel require bandwidth twice the required bandwidth over the conventional coherent channel. We propose a method to transmit without intersymbol interference  in a  bandwidth no larger than the bit rate. This is done by combining Nyquist or root-Nyquist pulses with a constant bias and using higher-order modulation formats. In fact, we can transmit with a bandwidth equal to that of coherent transmission. A trade-off between the required average optical power and the bandwidth is investigated. Depending on the  bandwidth required, the most power-efficient transmission is obtained by  the parametric linear pulse, the so-called ``better than Nyquist" pulse, or the root-raised cosine pulse.
\end{abstract}
\vspace{-5mm}
\begin{IEEEkeywords}\vspace{-3mm}Intensity-modulated direct-detection (IM/DD),  strictly bandlimited  signaling.
\end{IEEEkeywords}
\vspace{-6mm}
\section{Introduction}\vspace{-1mm}
\IEEEPARstart{T}{he growing} demand for high-speed data transmission systems has introduced
new design paradigms for optical communications. The need for low-complexity and cost-effective systems has motivated the usage of
affordable optical hardware (e.g., incoherent transmitters, optical
intensity modulators, multimode fibers, direct-detection receivers)
to design short-haul  optical fiber links (e.g., fiber to the home
and optical interconnects)\cite{4528765,5875697} and diffuse indoor wireless
optical links \cite{gfeller2005wireless,kahn2002wireless,hranilovic2005design}.
These devices impose three important constraints on the  signaling design. First, the transmitter
only modulates information on the instantaneous intensity
 of an optical carrier, contrary to conventional coherent
channels where the amplitude and phase of the carrier
can be used to send information \cite [Sec.~4.3] {proakis2001digital}. In the receiver, only the optical intensity
of the incoming signal will be detected  \cite{kahn2002wireless}. Due to these limitations,
the transmitted signal must be nonnegative. Such transmission is
called intensity modulation with direct detection (IM/DD). Second, the
peak and average optical power (i.e., the peak and average
of the transmitted signal in the electrical domain) must be below a certain threshold for eye- and skin-safety concerns \cite{kahn2002wireless}
and to avoid nonlinearities present in the devices \cite{westbergh200932,inan2009impact}.
 In conventional channels, such constraints are usually imposed on
the peak and average of the  squared electrical signal. Third, the
bandwidth is limited due to the impairments in the optoelectronic devices \cite{hranilovic2005design,4132995}
and other limitations (e.g.,
 modal dispersion in short-haul optical fiber links \cite{5223628} and  multipath distortion in diffuse indoor wireless
optical links\cite{kahn2002wireless}).
Consequently, the coherent modulation formats and pulse shaping methods designed for
conventional electrical channels  (i.e., with no nonnegativity
constraint on the transmitted signal) cannot be directly applied to IM/DD channels.

Pulse shaping for the purpose of reducing intersymbol
interference (ISI) in conventional channels has been previously investigated in \cite [Sec.~9] {proakis2001digital}, \cite{5055024,1089824,beaulieu2002better,1369626,chandan05,192386}.
Much research has been conducted on determining  upper and lower
bounds on the capacity of IM/DD channels considering power and bandwidth
limitations \cite{farid2010capacity,hranilovic2004capacity,farid2009capacity,lapidoth2009capacity,you2002capacity,you2002upper}.
In \cite{kahn2002wireless,16882,524287,780164,490239,HranilovicLattice,karout2010power,FSOIMDD}, the performance of various modulation formats in IM/DD channels
were studied using rectangular or other time-disjoint (i.e.,
infinite-bandwidth) pulses. 

Hranilovic in \cite{1577875} pioneered in investigating the problem of designing strictly bandlimited
pulses for IM/DD channels with nonnegative pulse-amplitude modulation (PAM)
schemes. He showed the existence of nonnegative bandlimited Nyquist pulses,  which can be used for ISI-free transmission over IM/DD channels, and evaluated the performance of  such pulses. He also showed that  any nonnegative root-Nyquist pulse must be time limited (i.e., infinite bandwidth). Hence receivers with matched filters are not suitable for Hranilovic's  signaling method. He concluded that  transmission is possible with a bandwidth twice the required bandwidth over the corresponding  conventional electrical channels. This work was extended to other Nyquist pulses that can introduce a  trade-off between bandwidth and average optical power in \cite{4132995,bhandari2007squared}.

In this paper, we present a new  signaling method for bandlimited
IM/DD channels, in which the
transmitted signal becomes nonnegative by the addition of a constant direct-current (DC) bias. This method provides us with two benefits: (i)
We can transmit ISI-free with a bandwidth equal to that of
coherent conventional channels, while benefiting from the reduced complexity
and cost of IM/DD system. (ii) We can implement the system using either Nyquist pulses with sampling receiver or root-Nyquist pulses with matched filter receiver. By being able to use a larger variety of pulses, the transmitted power can be reduced compared with known methods, which is advantageous in power-sensitive optical interconnects and indoor wireless optical links. We also evaluate the spectral efficiency
and optical power efficiency of binary and 4-PAM formats with Nyquist and root-Nyquist pulses for achieving a specific noise-free eye opening or a specific symbol-error-rate (SER).

The remainder of the paper is organized as follows. Section \ref{sec:SYSTEM-MODEL}
presents the system model. In Section \ref{sec:Pulse-shaping}, we define the Nyquist pulses that have been used extensively for conventional bandlimited channels,  as well as the ones that have been suggested for nonnegative bandlimited channels. In Section \ref{sec:RootPulse-shaping}, the root-Nyquist pulses used in this study are introduced. Section \ref{sec:Required-DC-bias} discusses
a method of computing the required DC bias for a general
pulse. Section \ref{sec:RESULTS} introduces the performance
measures and analyzes the performance of the system under different
scenarios. Finally, conclusions are drawn in Section \ref{sec:CONCLUSIONS} on the performance of the system.
\section{System Model\label{sec:SYSTEM-MODEL}}
In applications such as diffuse indoor wireless optical links and
short-haul  optical fiber communications, where inexpensive hardware is used,
IM/DD is often employed. In such systems, the data is modulated on the optical
intensity of the transmitted light using an optical intensity modulator such as a  laser diode or a light-emitting
diode. This optical intensity is proportional to the  transmitted electrical signal. As a result, the transmitted electrical
signal must be nonnegative. This is in contrast to conventional electrical channels, where the data is modulated on the amplitude and phase of the
carrier \cite [Sec.~4.3] {proakis2001digital}.  In the receiver, the direct-detection method is used in which the photodetector generates
an output which is proportional to the incident received instantaneous
power \cite{780164}. Another limitation,
which is considered for safety purposes, is a constraint on the peak and average
optical power, or equivalently, a constraint on the peak and average
of the signal in the electrical domain \cite{4132995,farid2010capacity,hranilovic2004capacity,lapidoth2009capacity,kahn2002wireless}. In this study, we consider
the IM/DD transmission system with a strict bandwidth limitation
and general $M$-level  modulation.

\begin{figure*} 
\psfrag{a}{\footnotesize $a_k$}
\psfrag{b}{\footnotesize $q(t)$}
\psfrag{c}{\footnotesize $\mu$}
\psfrag{d}{\footnotesize $I(t)$}
\psfrag{e}{Modulator}
\psfrag{f}{\footnotesize $x(t)$}
\psfrag{g}{\footnotesize $h(t)$}
\psfrag{h}{Detector}
\psfrag{i}{\footnotesize $n(t)$}
\psfrag{j}{\footnotesize $y(t)$}
\psfrag{k}{\footnotesize \hspace{1mm}$g(t)$}
\psfrag{l}{\footnotesize $r(t)$}
\psfrag{m}{\footnotesize $kT_\mathrm{s}$}
\psfrag{n}{\footnotesize $\hat{a}_{k}$}
\psfrag{o}{Electrical domain}
\psfrag{p}{Optical domain}
\psfrag{q}{Electrical domain}
 \begin{center}
\includegraphics[width=1\linewidth]{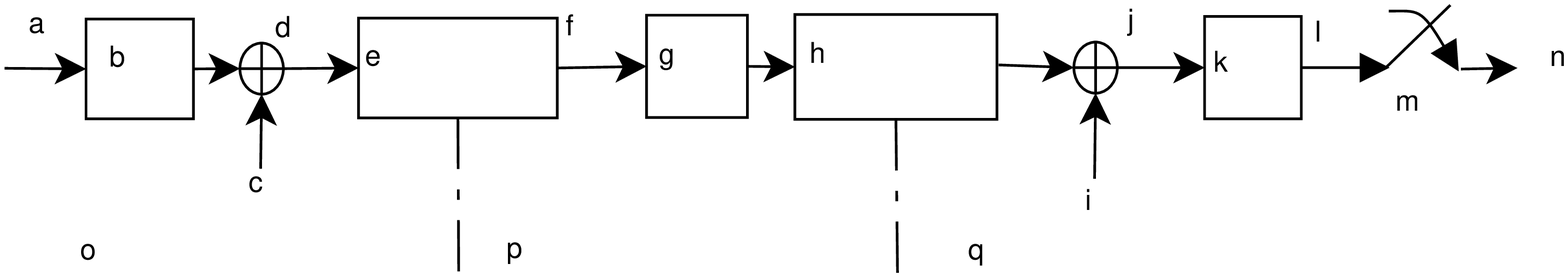}
\end{center}
\caption{Baseband system model, where $a_k$ is the  $k$-th input symbol, $q(t)$ is an arbitrary pulse, $\mu$ is the DC bias, $I(t)$ is the transmitted electrical signal, $x(t)$ is the optical intensity, $h(t)$ is the channel impulse response, $n(t)$ is the Gaussian noise, $g(t)$ is the impulse response of the receiver filter, and  $\hat{a}_{k}$ is an estimate of $a_k$.}
\label{fig:system}
\end{figure*}

Fig.~\ref{fig:system} represents the system model for an IM/DD optical
transmission system. It can be modeled as an
electrical baseband transmission system with additive white Gaussian noise (AWGN) and a nonnegativity
constraint  on the channel input \cite{4132995,kahn2002wireless,kahn2002experimental,gfeller2005wireless}.
We consider an ergodic source with independent and identically
distributed information symbols  $a_{k}\in\mathcal{C}$, where $k\in\mathbb{Z}$ is the discrete time instant, and $\mathcal{C}$ is a finite set of constellation points. Based on these symbols,  an electrical
signal $I(t)$ is generated.
 The optical intensity modulator converts
the electrical signal to an optical signal with optical carrier frequency $f_{\mathrm{c}}$ and random phase $\theta$, given by
$O(t)=\sqrt{2x(t)} cos\left(2\pi f_{\mathrm{c}} t+\theta \right)$,
where $x(t)$ is the intensity of the optical signal. This intensity is a linear function of $I(t)$ \cite{kahn2002wireless}, given by
\begin{equation}
x(t)=JI(t)=JA\left(\mu+\sum_{k=-\infty}^{\infty}a_{k}q(t-k\Ts)\right),\label{eq:TxSignal}\end{equation}
where $J$ is the laser conversion factor, $A$ is a scaling factor that can be adjusted depending on the desired transmitted power, $\mu$ is the required DC bias, $q(t)$ is an arbitrary  pulse, and  $\Ts$ is the symbol duration.

\begin{figure}
\psfrag{x(t)/A}{\footnotesize $x(t)/A$}
\psfrag{t}{\footnotesize \hspace{2mm}$t/\Ts$}

\includegraphics[width=1\textwidth]{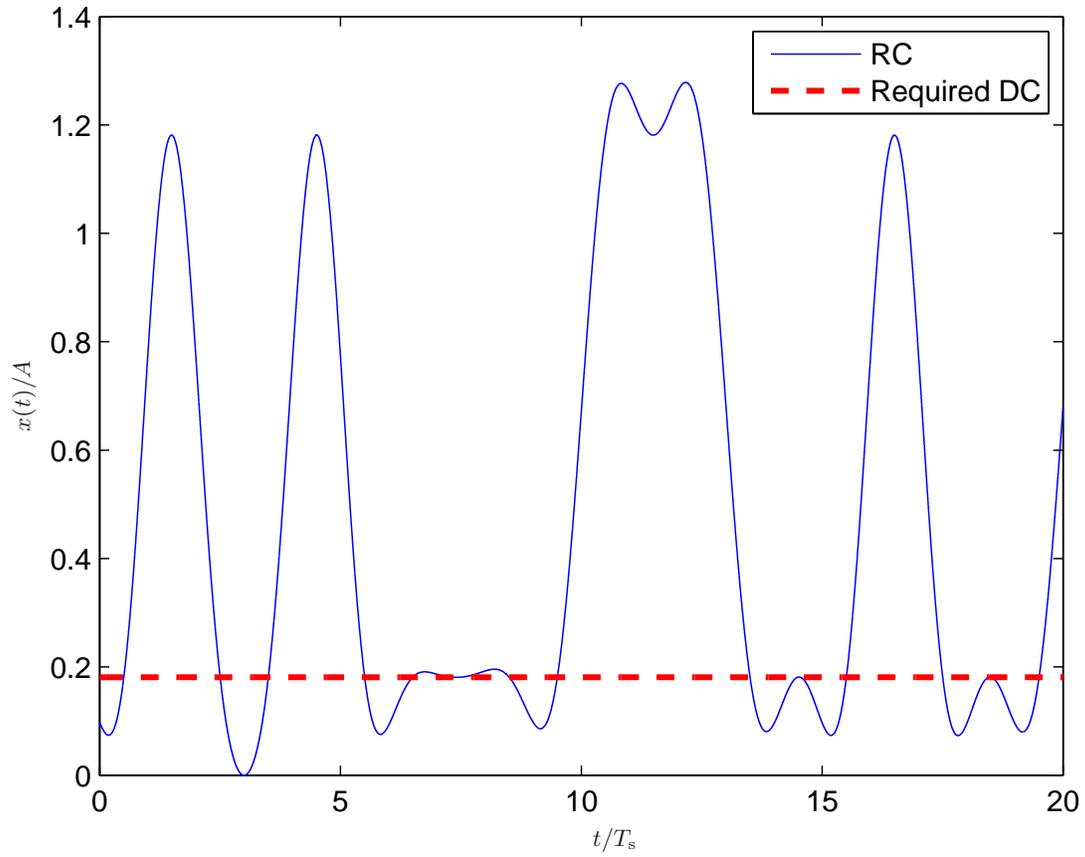}
%

\caption{\label{fig:TxSig_RC}The normalized transmitted signal $x(t)/A$ for $\mathcal{C}= \left\{0,1\right\}$ and
using an RC pulse with $\alpha=0.6$ as $q(t)$. It can be seen that without using the bias $\mu=0.184$ (see Fig.~\ref {fig:RequiredDC}), the RC pulse would create a signal $x(t)$ that can be negative.}
\end{figure}

Three requirements are placed on $x(t)$: it should be nonnegative, bandlimited, and ISI-free. The nonnegativity constraint, $x(t)\geq0$ for all  $t\in\mathbb{R}$,  is fulfilled by choosing $\mu$ in (\ref{eq:TxSignal}) sufficiently large, see Sec.~\ref{sec:Required-DC-bias}. This DC
bias is added equally to each symbol to maintain a  strictly bandlimited
 signal $x(t)$, in contrast to works like \cite{karout2010power,780164,HranilovicLattice} in which the bias is allowed to vary with time. The bandwidth constraint is fulfilled by choosing the pulse $q(t)$ such that
\begin{equation}
Q(\omega)=\intop_{-\infty}^{\infty}q(t)e^{-j\omega t}dt=0,\; |\omega|\geq 2\pi B,\label{eq:bandwidthLimitation}\end{equation} where $Q(\omega)$ denotes the Fourier transform of $q(t)$. The condition of ISI-free transmission, finally, is fulfilled by either choosing $q(t)$ as a Nyquist pulse, see Sec.~\ref{sec:Pulse-shaping}, when using a sampling receiver, or choosing $q(t)$ as a root-Nyquist pulse (also known as $\Ts$-orthogonal pulse), see Sec.~\ref{sec:RootPulse-shaping}, when using a matched filter in the receiver. Fig.~\ref{fig:TxSig_RC} illustrates an example of the transmitted intensity given by (\ref{eq:TxSignal}) where  $\mathcal{C}=\left\{ 0,1\right\} $.

Depending on the application, it is desirable to minimize  the average optical power or the peak optical power
\cite{4132995,hranilovic2004capacity,lapidoth2009capacity,kahn2002wireless,westbergh200932,inan2009impact}. The average optical power is
\begin{equation}
P_{\mathrm{opt}}=\frac{1}{\Ts}\intop_{0}^{\Ts}\mathbb{E}\left\{ x(t)\right\} dt,\nonumber\end{equation} where  $\mathbb{E}\left\{ \cdot\right\} $ denotes expectation, which for the definition of $x(t)$ in (\ref{eq:TxSignal}) yields
\begin{align}
P_{\mathrm{opt}} & =\frac{1}{\Ts}\intop_{0}^{\Ts}JA\left(\mu+\mathbb{E}\left\{ a_{k}\right\}\sum_{k=-\infty}^{\infty}q(t-k\Ts)\right)dt\nonumber\\
&=JA\left(\mu+\mathbb{E}\left\{ a_{k}\right\}\overline{q}\right),\label{eq:OpticalPower}\end{align}
where
\begin{equation}
\overline{q}=\frac{1}{\Ts}\intop_{-\infty}^{\infty}q(t)dt=\frac{Q(0)}{\Ts}.\label{eq:Overbar_q}\end{equation}
The peak optical power is
\begin{equation}
P_\mathrm{max} = \max x(t)= JA \left(\mu+\max \sum_{k=-\infty}^{\infty} a_k q(t-k\Ts) \right)
\label{pmax}
\end{equation}where the maximum is taken over all symbol sequences $\ldots, a_{-1}, a_0, a_1, a_2,\ldots$ and all times $t$.

The optical signal then propagates through the channel and is detected
and converted to the electrical signal  \cite{kahn2002wireless,hranilovic2004capacity}
\vspace{-5mm}
\begin{equation}
y(t)=Rh(t)\otimes x(t)+n(t),\nonumber\end{equation}
where $R$ is the responsivity of the photodetector, $\otimes$ is the convolution operator, $h(t)$ is the
channel impulse response, and $n(t)$ is the noise.
 In this study, the channel is considered to be flat in the bandwidth
of interest, i.e.,
$h(t)=H(0)\delta(t)$. Without loss of generality, we assume that $R=J=1$ \cite{kahn2002wireless} and $H(0)=1$. Since the thermal noise of the receiver and the shot
noise induced by ambient light  are two major noise sources in this setup, which are independent from the signal,
$n(t)$ can be modeled as a zero-mean AWGN
with double-sided power spectral density $N_0/2$ \cite{kahn2002wireless,audeh2002performance,lapidoth2009capacity,proakis2001digital}.
Although the input signal to the channel $x(t)$ must be nonnegative,
there is no such constraint on the received signal $y(t)$ \cite{farid2010capacity}.

The received signal passes through a filter with impulse response $g(t)$, resulting in\vspace{-3mm}
 \begin{equation}
r(t)=y(t)\otimes g(t),\label{eq:FilteredSignal}\end{equation}
which is then sampled at the symbol rate. In this paper, two scenarios are considered for the  receiver filter:

(i) Similarly to  \cite{1577875,4132995}, $y(t)$ can enter  a sampling receiver, which in this paper  is assumed
to have a rectangular  frequency response to limit the power of the noise in the
receiver, and is  given by
\begin{equation}
G(\omega)=\begin{cases}
\begin{array}{cc}
G(0) & |\omega|<2\pi B\\
0 & |\omega|\geq 2 \pi B\end{array}.\end{cases}\label{eq:G(f)}\end{equation}

(ii) According to our proposed method, $y(t)$ can enter a matched filter receiver with frequency response $G(\omega)=\zeta Q^{*}(\omega)$ where {$\left( \cdot \right)^*$} is the complex conjugate and $\zeta$ is an arbitrary scaling factor. This type of filter will limit the power of the noise, and can also result in ISI-free transmission if the pulses are root-Nyquist (see Sec.~\ref{sec:RootPulse-shaping}).

The system model introduced in this section is a generalization of the one in \cite{4132995}, which is obtained by considering $\mathcal{C}\subset\mathbb{R}^{+}$ and setting $\mu=0$ in (\ref{eq:TxSignal}). If $\mu=0$, the pulse $q(t)$ should be nonnegative to guarantee a  nonnegative signal $x(t)$. In our proposed system model, by introducing the bias $\mu$, the nonnegativity condition can be fulfilled for a wider selection of pulses $q(t)$ and constellation $\mathcal{C}\subset\mathbb{R}$.
\vspace{-5mm}
\section{Bandlimited Nyquist Pulses\label{sec:Pulse-shaping}}

In order to have ISI-free transmission with a sampling receiver,
the pulse $q(t)$ must satisfy the Nyquist criterion \cite{5055024}.
In other words, for any $k\in\mathbb{Z}$ \cite [Eq.~(9.2-11)] {proakis2001digital},
\begin{equation}
q(k\Ts)=\begin{cases}
q(0), & k=0,\\
0, & k\neq0.\end{cases}\label{eq:NyquistInTime}\end{equation}
The most popular Nyquist pulses are the classical ``sinc'' pulse, defined as $\sinc(x)= \sin(\pi x)/(\pi x)$, and the raised-cosine (RC) pulse \cite [Sec.~9.2]{proakis2001digital}. Many other Nyquist pulses have been proposed recently for the conventional channel; see \cite{alexandru11,assimonis11} and references therein.

In this paper, we evaluate some of these pulses, defined in Table~\ref{pulse-def}, for IM/DD transmission. Our selected pulses are the RC pulse, the so-called ``better than Nyquist''  (BTN) pulse \cite{beaulieu2002better}, which in \cite{1369626} was referred to as the parametric exponential pulse, the parametric linear  (PL) pulse of first order \cite{1369626}, and one of the polynomial (Poly) pulses in \cite{chandan05}. Their bandwidth can be adjusted via the parameter $0 \le \alpha \le 1$ such that their
lowpass bandwidth is $B=(1+\alpha)/(2\Ts)$. Since these pulses may be negative, they must be used in a system with $\mu>0$. We denote these four pulses as \emph{regular} Nyquist pulses.


Another option is to use \emph{nonnegative} Nyquist pulses, which satisfy all the three aforementioned constraints. As a result, in (\ref{eq:TxSignal}),
$\mu=0$ and $q(t)\geq0$ for all $t\in\mathbb{R}$.
In \cite{4132995}, it has been shown that  pulses that satisfy these requirements
must be the square of a general Nyquist pulse. This will result in having pulses with bandwidth twice that of the original Nyquist pulses. Three pulses that satisfy these constraints were introduced in \cite{4132995}, and we use them in our study for compatibility with previous works: squared sinc (S2), squared RC (SRC), and squared double-jump (SDJ), also defined in Table~\ref{pulse-def}. Their low-pass bandwidth is $B=1/\Ts$ for S2 and $B=(1+\alpha)/\Ts$ for SRC and SDJ, where $0\le\alpha\le 1$.

%
%

\begin{figure}
\psfrag{x(t)/A}{\footnotesize $x(t)/A$}
\psfrag{t}{\footnotesize \hspace{2mm}$t/\Ts$}
\includegraphics[width=1\textwidth]{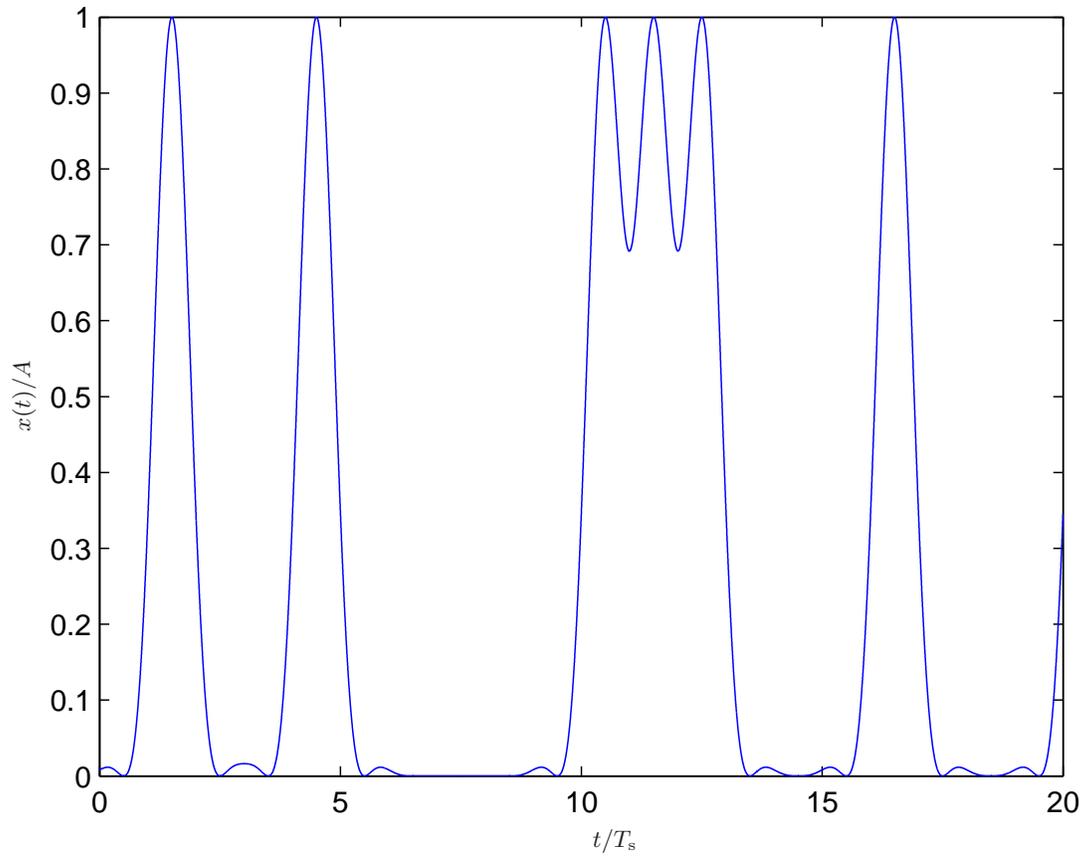}

\caption{\label{fig:TxSig_SquareRC}The normalized transmitted signal $x(t)/A$
 for $\mathcal{C}= \left\{0,1\right\}$ and using an SRC pulse with $\alpha=0.6$ as $q(t)$. In
this case, the required DC  $\mu$ is zero.}
%
\end{figure}

Figs. \ref{fig:TxSig_RC} and \ref{fig:TxSig_SquareRC} depict the
normalized transmitted signal $x(t)/A$ using the RC and SRC pulses, respectively, assuming $\mathcal{C}= \left\{0,1\right\}$.
The most important parameters of the pulses are summarized in Table \ref{tab:All_Info_Nyq}.

\vspace{-5mm}
\section{Bandlimited Root-Nyquist Pulses\label{sec:RootPulse-shaping}}
 ISI-free transmission is achieved with the pulses in Sec.~\ref{sec:Pulse-shaping} as long as the input of the sampling unit satisfies the Nyquist criterion given in (\ref{eq:NyquistInTime}). In addition to the method of using a Nyquist pulse in the transmitter and a rectangular filter \eqref{eq:G(f)} in the receiver, other scenarios can be designed that generate Nyquist pulses at the input $r(t)$ of the sampling unit. In one of these methods, the transmitted pulse is a   root-Nyquist pulse, and the receiver contains a filter matched to the transmitted pulse \cite [Sec.~5.1] {proakis2001digital}. Consequently, the output of the matched filter will be ISI-free if for any integer $k$ \vspace{-2mm} \begin{equation}
\intop_{-\infty}^{\infty}q(t)q(t-k\Ts)dt=\begin{cases}
E_{{q}} & k=0\\
0 & k\neq0
\end{cases}
\label{eq:Root_Nyquist Criteria},\end{equation}
where
$\Eq=\intop_{-\infty}^{\infty}q^{2}(t)dt.$
Tables \ref{pulse-def} and \ref{tab:All_Info_Nyq} also includes two root-Nyquist pulses that have been previously
used for conventional coherent channels, where again $0\leq \alpha \leq 1$.
These are the root raised cosine (RRC) pulse and the first-order Xia pulse \cite{xia1997family}. Both have  the lowpass bandwidth $B=(1+\alpha)/(2\Ts)$.



Although the output of the matched filter for both the first order Xia pulse and the RRC pulse are similar ($r(t)$ consists of RC pulses in both cases), the RRC is symmetric in time, whereas the Xia pulse has more energy in the precursor (i.e., the part of the pulse before the peak) \cite{tan2004transmission}. Moreover, the maximum of Xia pulse does not happen at the origin. The important point with the Xia pulse is that it is both a Nyquist and a root-Nyquist pulse.

In contrast to Nyquist pulses, from which nonnegative Nyquist pulses can be generated by squaring the original pulse (see Sec.~\ref{sec:Pulse-shaping}), the square of a root-Nyquist pulse is not root-Nyquist anymore. Moreover, \cite{4132995} has proven that there is no nonnegative root-Nyquist pulse with strictly limited bandwidth.


\vspace{-5mm}
\section{Required DC Bias \label{sec:Required-DC-bias}}
Our goal is to find the lowest $\mu$ that guarantees the nonnegativity of $x(t)$. From (\ref{eq:TxSignal}) and $x(t)\geq0$, the smallest required DC bias is
\vspace{-1mm}
\begin{align}
\mu & =-\min_{\forall a,-\infty< t < \infty}\sum_{k=-\infty}^{\infty}a_{k}q(t-k\Ts) \label{eq:DC1}\\
 &=-\min_{\forall a,-\infty< t < \infty}\sum_{k=-\infty}^{\infty}\left[\left(a_{k}-L\right)q(t-k\Ts)+Lq(t-k\Ts)\right] \label{eq:DC2}\end{align}
where $L=(\hat{a}+\check{a})/2$, $\hat{a}=\max_{a\in \mathcal{C}}a$, and $\check{a}=\min_{a\in \mathcal{C}}a$.
The notation $\forall a$ in \eqref{eq:DC1} and \eqref{eq:DC2} means that the minimization should be over all $a_k \in \mathcal{C}$ where $k = \ldots,-1,0,1,2,\ldots${} Going from (\ref{eq:DC1}) to (\ref{eq:DC2}), we created a factor ($a_{k}-L$) which is a function of $a_{k}$ and symmetric with respect to zero. As a result,  the minimum of the first term in (\ref{eq:DC2}) occurs if, for all $k$, either $a_{k}=\hat{a}$ and $q(t-k\Ts)<0$ or $a_{k}=\check{a}$ and $q(t-k\Ts)>0$. In both cases, due to the fact that the factor $\hat{a}-L=-(\check{a}-L)$,
\vspace{-1mm}
\begin{align}
 \mu &=\max_{0\leq t< \Ts}\left[\left(\hat{a}-L\right)\sum_{k=-\infty}^{\infty}\left|q(t-k\Ts)\right|-L\sum_{k=-\infty}^{\infty}q(t-k\Ts)\right].\label{eq:DC}\end{align}
The reason why (\ref{eq:DC}) is minimized over $0\leq t< \Ts$ is
that $\sum_{k=-\infty}^{\infty}q(t-k\Ts)$ and   $\sum_{k=-\infty}^{\infty}|q(t-k\Ts)|$ are periodic functions
with period equal to $\Ts$. Since for all pulses defined in Sec.~\ref{sec:Pulse-shaping} and \ref{sec:RootPulse-shaping},  $q(t)$ rescales with $\Ts$ as $q(t)=v(t/\Ts)$ for some function $v(t)$, then  $\mu$ is independent of $\Ts$.

To simplify (\ref{eq:DC}),   Lemma~\ref{LemFS} and  Corollary~\ref{CorrAvval} will be helpful, since they prove that the second term in (\ref{eq:DC}) does not change over time.

\begin{lemma} \label{LemFS} For an arbitrary pulse $q(t)$,
\[
\sum_{k=-\infty}^{\infty}q(t-k\Ts)=\frac{1}{\Ts}\sum_{n=-\infty}^{\infty}Q\left(\frac{2\pi n}{\Ts}\right)e^{\frac{j2\pi nt}{\Ts}}.\]

\end{lemma}

\begin{IEEEproof} Since
$f(t)=\sum_{k=-\infty}^{\infty}q(t-k\Ts)$ is a periodic function with period $\Ts$, it can be expanded as a Fourier series. Its  Fourier series coefficients
are
\begin{align}
C_{n} & =\frac{1}{\Ts}\intop_{-\Ts/2}^{\Ts/2}f(t)e^{-\frac{j2\pi nt}{\Ts}}dt\nonumber \\
&=\frac{1}{\Ts}\intop_{-\Ts/2}^{\Ts/2}\sum_{k=-\infty}^{\infty}q(t-k\Ts)e^{-\frac{j2\pi nt}{\Ts}}dt.\label{eq:Fourier Coeffs}\end{align}
Since both $n$ and $k$ are integers, $e^{j2\pi nk}=1$. As a result,
(\ref{eq:Fourier Coeffs}) can be written as
\begin{align}
C_{n} & =\frac{1}{\Ts}\intop_{-\Ts/2}^{\Ts/2}\sum_{k=-\infty}^{\infty}q(t-k\Ts)e^{-\frac{j2\pi n}{\Ts}(t-k\Ts)}dt\nonumber \\
& =\frac{1}{\Ts}\intop_{-\infty}^{\infty}q(t)e^{-\frac{j2\pi nt}{\Ts}}dt=\frac{1}{\Ts}Q\left(\frac{2\pi n}{\Ts}\right).\nonumber \end{align}
Hence, \vspace{-3mm}
\begin{equation}
f(t)=\sum_{n=-\infty}^{\infty}C_{n}e^{\frac{j2\pi nt}{\Ts}}=\frac{1}{\Ts}\sum_{n=-\infty}^{\infty}Q\left(\frac{2\pi n}{\Ts}\right)e^{\frac{j2\pi nt}{\Ts}},\label{eq:bandlimited}\end{equation}
which proves the lemma.\end{IEEEproof}

%

%

The usefulness of this lemma follows from the fact that for bandlimited pulses $q(t)$, (\ref{eq:bandlimited}) is reduced to a finite number of terms. As a special case, we have the following corollary.

\begin{corollary}If $q(t)$ is a bandlimited pulse defined in (\ref{eq:bandwidthLimitation}), where $B\Ts\leq 1$,
then (\ref{eq:bandlimited}) can be written as \vspace{-3mm}
\begin{equation}
f(t)=\sum_{k=-\infty}^{\infty}q(t-k\Ts)=\frac{1}{\Ts}Q(0).\label{eq:coroll1}\end{equation}In other words, for such $q(t)$, this sum is not a function of
time.
\label{CorrAvval}\end{corollary}
\begin{IEEEproof}
Since $B\Ts\leq 1$, the sum in (\ref{eq:bandlimited}) has only one nonzero term (i.e.,  $Q(0)$ can be nonzero whereas  $Q(2\pi n /\Ts)=0$ for all $n\neq 0$ due to  (\ref{eq:bandwidthLimitation})).\end{IEEEproof}

As a result of Corollary \ref{CorrAvval}, (\ref{eq:DC}) for the regular Nyquist pulses  and root-Nyquist pulses considered in Sec.~\ref{sec:Pulse-shaping} and \ref{sec:RootPulse-shaping} (but not SRC and SDJ) can be written as
\begin{equation}
\mu=\left(\hat{a}-L\right)\max_{0\leq t< \Ts}\sum_{k=-\infty}^{\infty}\left|q(t-k\Ts)\right|-L\frac{Q(0)}{\Ts},\label{eq:DC_New}\end{equation}
where $Q(0)=\overline{q}\Ts$ for all pulses, see (\ref{eq:Overbar_q}). It appears that solving the summation in (\ref{eq:DC_New}) is impossible analytically even for simple pulses.








\begin{theorem} For bandlimited pulses  where $B\Ts\leq 1$, the
transmitted signal (\ref{eq:TxSignal}) is unchanged if all constellation points in $\mathcal{C}$ are shifted by a constant offset.
\label{CorrSevom}\end{theorem}
\begin{IEEEproof} Since the chosen pulse has limited bandwidth given by (\ref{eq:bandwidthLimitation}), using (\ref{eq:coroll1}) given in Corollary \ref{CorrAvval},  the transmitted signal (\ref{eq:TxSignal}) can be written as
\vspace{-1mm}
\begin{align}
x(t)=A\left(\mu+\sum_{k=-\infty}^{\infty}\left(a_{k}-L+L\right)q\left(t-k\Ts\right)\right)\nonumber\\ = A\left(\mu+\sum_{k=-\infty}^{\infty}\left(a_{k}-L\right)q\left(t-k\Ts\right)+L\frac{Q(0)}{\Ts}\right)\label{eq:alaki}.\end{align}
Substituting the required bias given by (\ref{eq:DC_New}), (\ref{eq:alaki}) can  be written as
\begin{equation}
x(t)=A\Bigg(\left(\hat{a}-L\right)\max_{0\leq t< \Ts}\left[\sum_{i=-\infty}^{\infty}\left|q(t-i\Ts)\right|\right]+\sum_{k=-\infty}^{\infty}\left(a_{k}-L\right)q\left(t-k\Ts\right)\Bigg).\label{eq:symmetric}\end{equation}
It can be seen that (\ref{eq:symmetric}) only depends on symbols through  $\hat{a} - L$ and $a_{k}-L$. Both terms are independent of the constellation offset.\end{IEEEproof}

Theorem \ref{CorrSevom} shows that for narrow-band pulses defined in (\ref{eq:bandwidthLimitation}), the constellation offset does not have an effect on the performance. This result which holds for intensity modulated channels (with nonnegative transmitted signal requirement) is in contrast to the standard result for conventional channels. For instance, binary phase-shift keying (BPSK) and on-off keying (OOK) are equivalent in this IM/DD system, whereas BPSK is 3 dB better over the conventional AWGN channel \cite[Sec.~5]{proakis2001digital}.

Fig.~\ref{fig:RequiredDC} illustrates
the required DC bias (\ref{eq:DC_New}) for various pulses considering any nonnegative $M$-PAM  constellation ($\mathcal{C}=\left\{ 0,1,...,M-1\right\} $). In case of Nyquist pulses, due to the fact that by increasing $\alpha$, the ripples of the pulses decrease, the required DC bias decreases as well. It can be seen that the Poly and RC pulses always require more DC bias than other Nyquist pulses. The PL and BTN pulses require approximately  the same DC bias.  The BTN pulse requires slightly less DC bias in  $0.250\le \alpha \le 0.256$, $0.333\le \alpha \le 0.363$, and $0.500 \le \alpha \le 0.610$, while the  PL is better for all other roll-off factors in the range $0 < \alpha < 1$.

The RRC pulse has a different behavior. For $0< \alpha \le 0.420$, similar to Nyquist pulses, by increasing the roll-off factor, the required DC bias decreases, and is approximately equal to the required DC bias for BTN and PL. However, when $0.420\le \alpha < 1$, the required DC bias starts to fluctuate slightly around $\mu=0.25\hat{a}$ and the  minimum happens for $\alpha =0.715$. The reason for this behavior is that in RRC, the peak is a function of $\alpha$, see Table~\ref{pulse-def}. As a result, by increasing the roll-off factor, there will be a compromise between the reduction in the sidelobe amplitude and the increase in peak amplitude. For small values of $\alpha$, the sidelobe reduction is more significant than the peak increase, and as a result, the required DC bias decreases.
The Xia pulse always requires the largest DC bias. For $0< \alpha \le 0.730$, similar to other pulses, by increasing the roll-off factor, the required DC bias for Xia pulses decreases. However, when $0.730\le \alpha < 1$, the required DC bias starts to fluctuate slightly and starts to approach the required DC for RRC.



The expression for $\mu$ given in (\ref{eq:DC}) illustrates the reason why the  double-jump and sinc pulses are not considered in  Sec.~\ref{sec:Pulse-shaping}. These pulses decay as $1/|t|$. As a result, the summation in (\ref{eq:DC}) does not converge to a finite value. Hence, they require an infinite amount of DC bias to be  nonnegative.
\vspace{-7mm}
\section{Analysis and Results\label{sec:RESULTS}}
\vspace{-1.5mm}
\subsection{\vspace{-1.5mm}Received Sequence for Sampling Receiver\label{subsec:RxSequence_Nyq}}

Considering the  assumptions mentioned in Sec.~\ref{sec:SYSTEM-MODEL}, the received signal (\ref{eq:FilteredSignal})
is \vspace{-3mm}
\begin{align}
r(t) &=\left(x(t)+n(t)\right)\otimes g(t) \nonumber\\
 &=A\left(\mu+\sum_{k=-\infty}^{\infty}a_{k}q(t-k\Ts)\right)\otimes g(t)+z(t) \nonumber\\
 & =AG(0)\left[\mu+\sum_{k=-\infty}^{\infty}a_{k}q(t-k\Ts)\right]+z(t),\label{eq:receivedSignal}\end{align}
where (\ref{eq:receivedSignal}) holds since $g(t)$ has a flat frequency response given by (\ref{eq:G(f)}) over the bandwidth of $q(t)$  given by (\ref{eq:bandwidthLimitation}); Therefore, the
 convolution has no effect on  $x(t)$. The noise
at the output of the receiver filter, which is given by  $z(t)=n(t)\otimes g(t)$, is zero mean additive
white Gaussian with variance $\sigma_{z}^2=G(0)^2N_{0}B$.

 Applying the Nyquist criterion given in (\ref{eq:NyquistInTime}) to the sampled version
of (\ref{eq:receivedSignal}), we can write the $i$-th filtered sample
as \vspace{-5mm} \begin{equation}
r(i\Ts)=AG(0)\left[\mu+a_{i}q(0)\right]+z(i\Ts).\label{eq:RecSamplesEQ}\end{equation}
for any constellation $\mathcal{C}$. The received waveform $r(t)$, for several Nyquist pulses, is shown in Fig.~\ref{eye-diagrams}, in the form of eye diagrams in a noise-free setting ($z(t) = 0$). As expected, the output samples $r(i \Ts)$ are ISI-free.\vspace{-3mm}
\subsection{\vspace{-2mm}Received Sequence for Matched Filter Receiver\label{subsec:RxSequence_MF}}

Similar to Sec.~\ref{subsec:RxSequence_Nyq}, the received signal will be \vspace{-3mm}
\begin{align} \vspace{-3mm}
r(t) & =\left(x(t)+n(t)\right)\otimes g(t) \nonumber \\
& =A\left(\mu+\sum_{k=-\infty}^{\infty}a_{k}q(t-k\Ts)\right)\otimes \zeta q(-t)+u(t) \nonumber \\
& =A\zeta \Big(\mu\intop_{-\infty}^{\infty}q(-t)dt+\sum_{k=-\infty}^{\infty}a_{k}\intop_{-\infty}^{\infty}q(\tau-k\Ts)q(\tau-t)d\tau\Big)+u(t) \nonumber \\
& =A\zeta \left(\mu Q(0)+\sum_{k=-\infty}^{\infty}a_{k}\intop_{-\infty}^{\infty}q(\tau)q(\tau-t+k\Ts)d\tau\right)+u(t)\label{eq:MF_Output}\end{align}
where $u(t)$ is zero mean additive white Gaussian noise with variance $\sigma_{u}^{2}=\zeta^2N_{0}\Eq/2$.
Applying the root-Nyquist criterion given in (\ref{eq:Root_Nyquist Criteria}) to the sampled version
of (\ref{eq:MF_Output}), the $i$-th filtered sample will be, for any constellation $\mathcal{C}$, \vspace{-3mm}
\begin{equation}
r(i\Ts)=A\zeta \left(\mu Q(0)+a_{i}\Eq\right)+u(i\Ts).\label{eq:Rec_MF}\end{equation}

\vspace{-6mm}
\subsection{\vspace{-2mm}Comparison Between Pulses}


As mentioned in Sec.~\ref{sec:SYSTEM-MODEL}, it may be desirable to minimize the average or peak optical power. The next theorem shows that these two criteria are equivalent for narrow-band pulses ($B\Ts < 1$) and symmetric constellations ($\mathbb{E}\{a_k\} = L$).

\begin{theorem} \label{th:pmax}
If $B\Ts < 1$ and $\mathbb{E}\{a_k\} = L$, then $P_\mathrm{max} = 2 P_\mathrm{opt}$.
\end{theorem}
\begin{IEEEproof}
From \eqref{pmax} and Corollary \ref{CorrAvval},
\begin{align*}
P_{\mathrm{max}} &=JA\left(\mu+\max_{\forall a,-\infty< t < \infty}\sum_{k=-\infty}^{\infty}\left[\left(a_{k}-L\right)q(t-k\Ts)+Lq(t-k\Ts)\right]\right)\\
&=JA\left(\mu+\max_{\forall a,-\infty< t < \infty}\left[\sum_{k=-\infty}^{\infty}\left(a_{k}-L\right)q(t-k\Ts)+\frac{LQ(0)}{\Ts}\right]\right).
\end{align*}
In analogy to \eqref{eq:DC_New}, the maximum is
\begin{align*}
P_{\mathrm{max}} &=JA\left(\mu+(\hat{a}-L)\max_{0\leq t<\Ts} \sum_{k=-\infty}^{\infty}|q(t-k\Ts)|+\frac{LQ(0)}{\Ts}\right) =JA\left(2\mu+2\frac{LQ(0)}{\Ts}\right)
\end{align*}
which compared with \eqref{eq:OpticalPower} completes the proof.
\end{IEEEproof}

To compare the optical
power of various pulses, a criterion called optical power gain
is used, which is defined as\cite{4132995}  \vspace{-4mm}
\begin{equation}
\Upsilon=10\log_{10}\left(\frac{P_{\mathrm{opt}}^{\mathrm{ref}}}{P_{\mathrm{opt}}}\right),\nonumber\end{equation}
where $P_{\mathrm{opt}}^{\mathrm{ref}}$ is the average optical power for a reference system. (According to Theorem~\ref{th:pmax}, $\Upsilon$ would be the same if defined in terms of $P_\mathrm{max}$, for all pulses in our study except SRC and SDJ.) Similarly to \cite{1577875}, this reference is chosen  to be the S2 pulse with OOK modulation and sampling receiver, for which no bias is needed. Using (\ref{eq:OpticalPower}), $P_{\mathrm{opt}}^{\mathrm{ref}}=A_{\mathrm{ref}}\mathbb{E}_{\mathrm{ref}}\left\{a_k\right\}$ and
\begin{equation}
\Upsilon=10\log_{10}\left(\frac{A_{\mathrm{ref}}\mathbb{E}_{\mathrm{ref}}\left\{a_k\right\}}{A\left(\mu+\mathbb{E}\left\{a_k\right\}\overline{q}\right)}\right)\label{eq:OptGainGeneral}\end{equation}
where $A_{\mathrm{ref}}$ and $\mathbb{E}_{\mathrm{ref}}\left\{a_{k}\right\}$ are the scaling factor and the symbol average for the reference system, respectively. 
Defining\vspace{-1mm}
\begin{equation}
\Delta a=\min_{a,a'\in\mathcal{C},a\neq a'}\left|a-a'\right|\label{eq:MinDistance}\end{equation}
as the minimum distance between  any two constellation points $a$ and $a'$, $\mathbb{E}_{\mathrm{ref}}\left\{a_k\right\}=\Delta a_{\mathrm{ref}}/2$, where $\Delta a_{\mathrm{ref}}$ is the minimum distance for the reference system. The expressions in (\ref{eq:OptGainGeneral}) and (\ref{eq:MinDistance}) hold in general for all finite set of constellation points $\mathcal{C}$.







\vspace{-2mm}
Initially, we compare the pulses in a noise-free setting. For any Nyquist pulse with a sampling receiver, the minimum eye opening after filtering is given by (\ref{eq:RecSamplesEQ}) as \vspace{-3mm}
\begin{equation}
\min_{ a,a'\in \mathcal{C}, a\neq a'} \left|AG(0)\left(\mu+aq(0)\right)-AG(0)\left(\mu+a'q(0)\right)\right|=AG(0)\Delta a q(0). \label{eq:MinEye_Sample}
\end{equation}
\vspace{-3mm}As a result, to have equal eye opening we require
${A_{\mathrm{ref}}}/{A}={\Delta a q(0)}/{\Delta a_{\mathrm{ref}}}$,
which substituted into \eqref{eq:OptGainGeneral} yields
\begin{equation}
\Upsilon=10\log_{10}\left(\frac{\Delta a q(0)}{\mu+\mathbb{E}\left\{a_k\right\}\overline{q}}\right).\label{eq:OptGainEye}\end{equation}

Fig.~\ref{fig:OptPower_Eye} demonstrates the comparison of
the optical power gain for various pulses defined in Sec.~\ref{sec:Pulse-shaping} for both OOK and 4-PAM  formats, where the signals are scaled to have equal eye opening. The S2 pulse with OOK modulation,
which is used as a baseline for comparison, is shown in the figure
with an arrow. The results for SRC and SDJ have been derived before in \cite[Fig.~4]{4132995}, whereas the results for other pulses are novel, where $T_{\mathrm{b}}=\Ts/\log_{2}M$ is the bit rate. OOK is chosen rather than BPSK for compatibility with \cite{4132995}, although these binary formats are entirely  equivalent for $BT_{\mathrm{b}}\leq 1$, as shown in Theorem \ref{CorrSevom}.
In these examples, we use $\Delta a = \Delta a_\mathrm{ref}$; however, rescaling the considered constellation $\mathcal{C}$ would not change the results, as it would affect the numerator and denominator of \eqref{eq:OptGainEye} equally.
For the nonnegative pulses in Sec.~\ref{sec:Pulse-shaping} (i.e., SRC and SDJ) with OOK, where $\mu=0$, by increasing the bandwidth, the optical power gain, which depends on $\alpha$ through its dependence on $\overline{q}$, increases since  $\overline{q}$ decreases.
The results in Fig.~\ref{fig:OptPower_Eye} are consistent with \cite[Fig.~4]{4132995}, where the same nonnegative pulses were presented. It can be seen that when the regular Nyquist pulses (RC, BTN, PL, and Poly) are used, and the nonnegativity constraint is satisfied by adding a DC bias,
 transmission is possible over a much narrower bandwidth.
However, since the DC bias consumes energy and does not carry information, the optical power gain will be reduced.

\begin{figure}
\psfrag{alpha}{\footnotesize $\alpha$}
\psfrag{mu}{\footnotesize $\mu/\hat{a}$}
\includegraphics[width=1\textwidth]{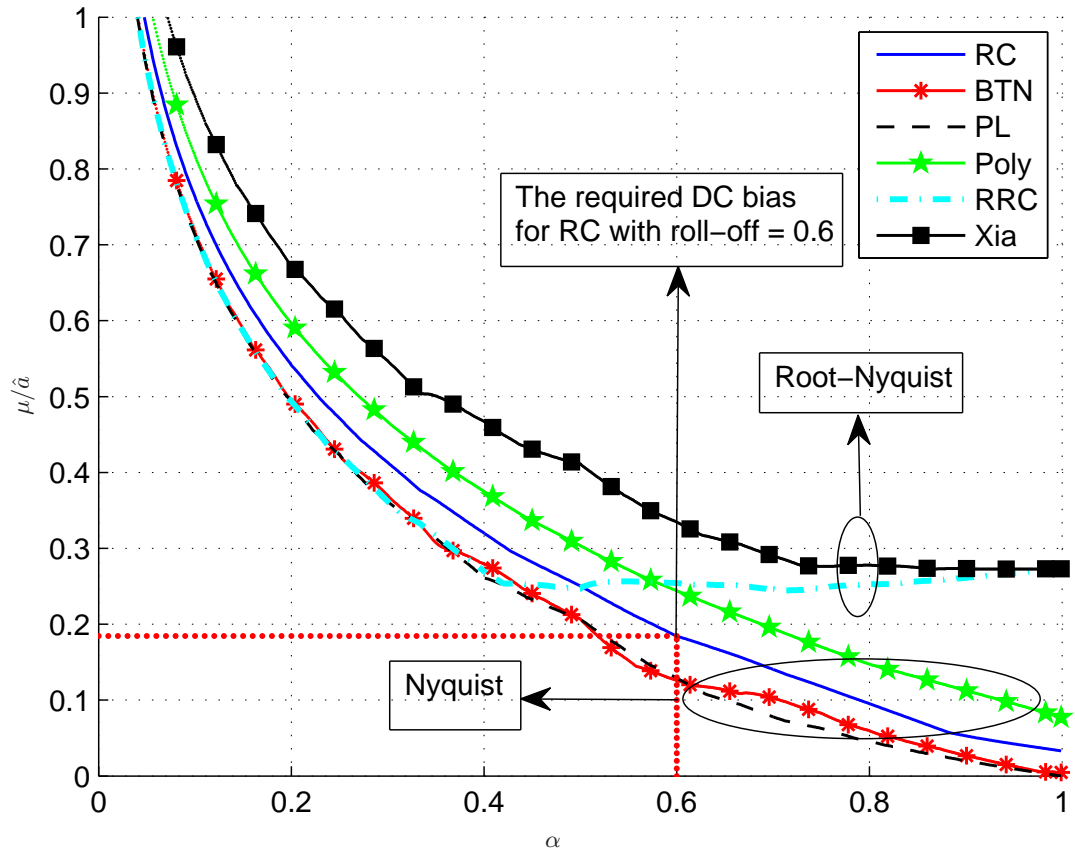}

\caption{\label{fig:RequiredDC}The  normalized minimum DC bias $\mu/\hat{a}$ vs. roll-off factor $\alpha$ for a variety of pulses and $M$-PAM. The dotted line represents the required bias for the RC pulse at $\alpha=0.6$, see Fig.~\ref{fig:TxSig_RC}.}
%
\end{figure}

\begin{figure}
\begin{center}
\psfrag{Time (t/T)}[t][t]{\footnotesize $t/\Ts$}
\psfrag{amplitude}{\footnotesize $r(t)$}
\begin{tabular}{cc}
{\includegraphics[width=7cm]{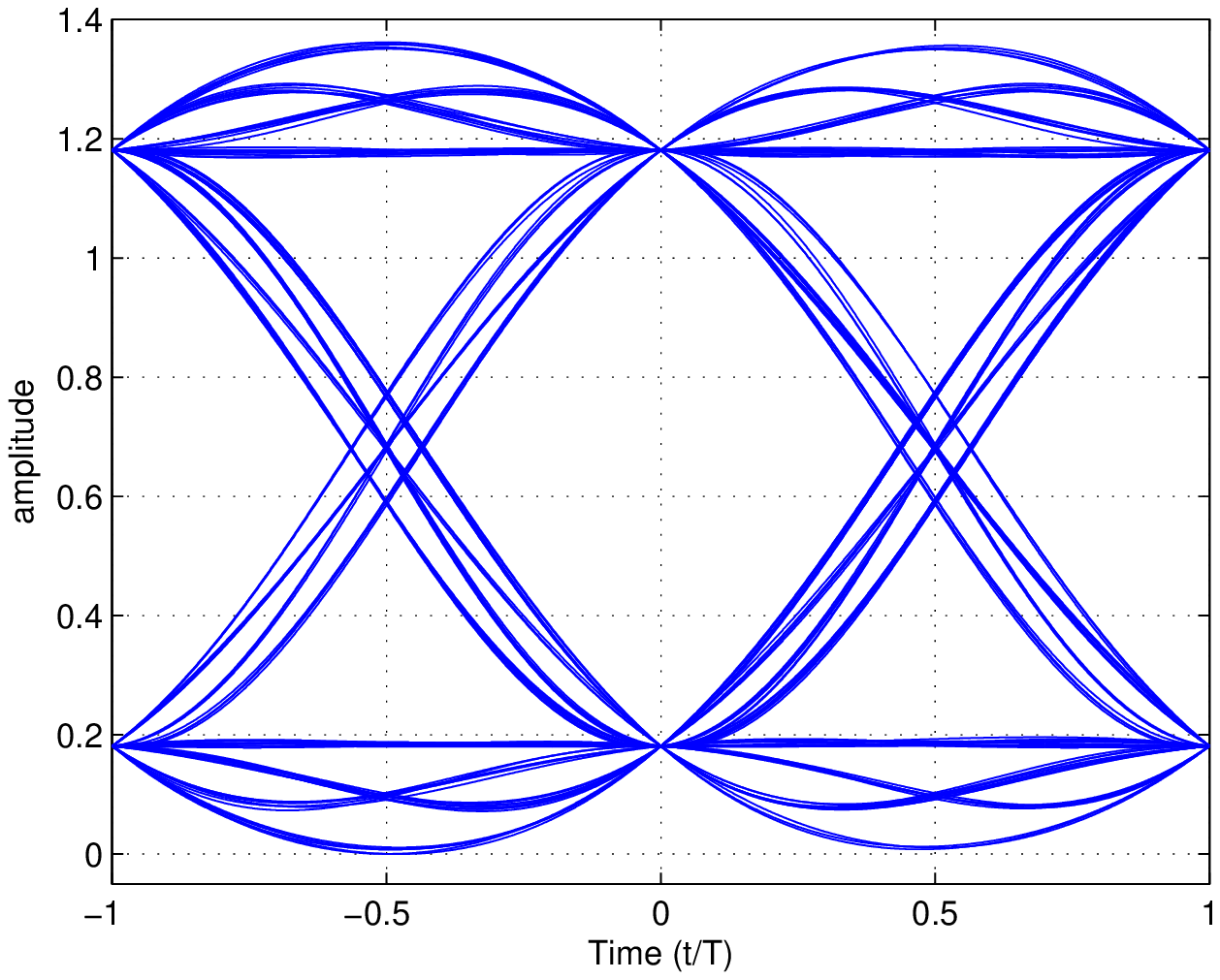}} &
{\includegraphics[width=7cm]{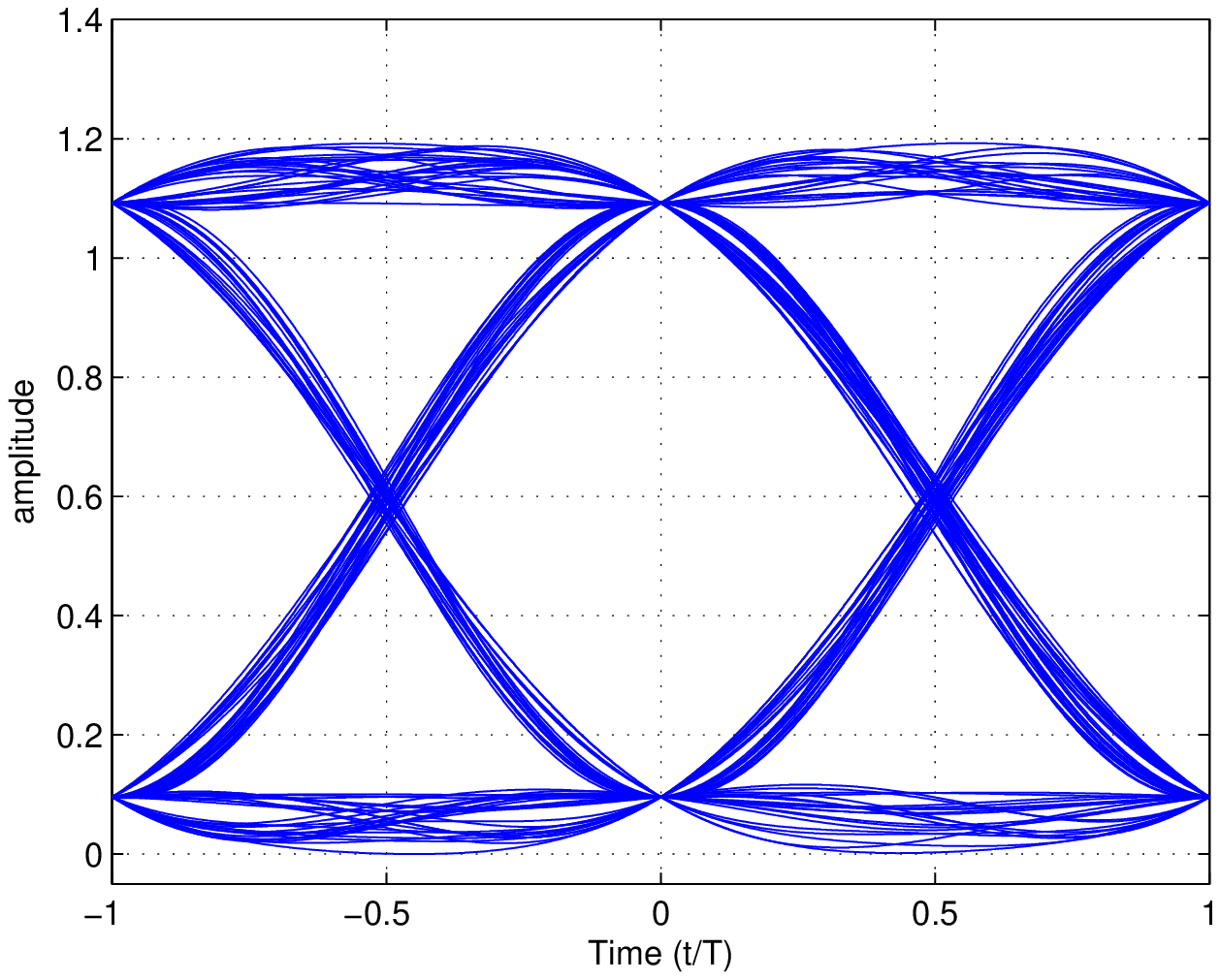}} \\[-2ex]
\small (a) & \small (b) \\
{\includegraphics[width=7cm]{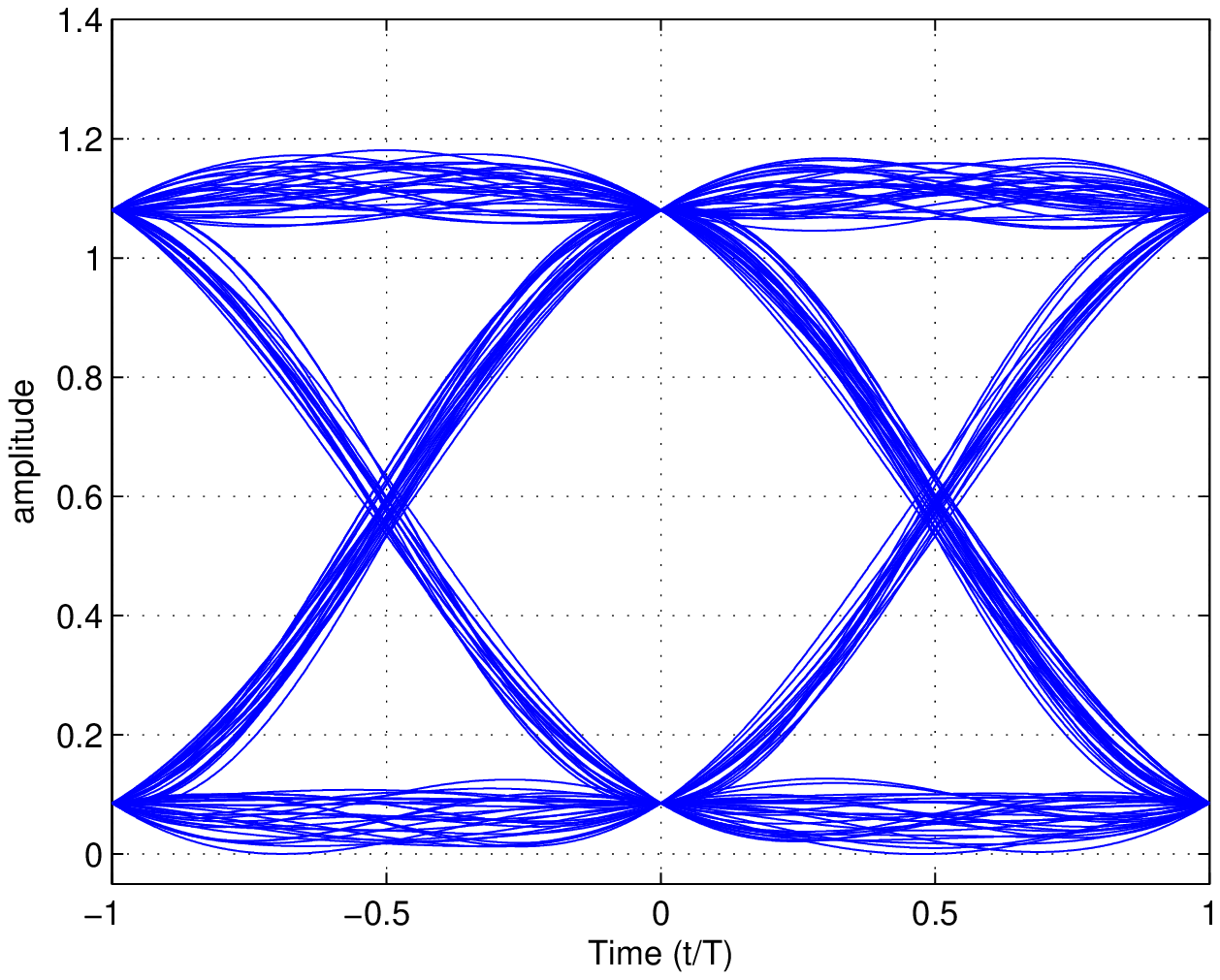}} &
{\includegraphics[width=7cm]{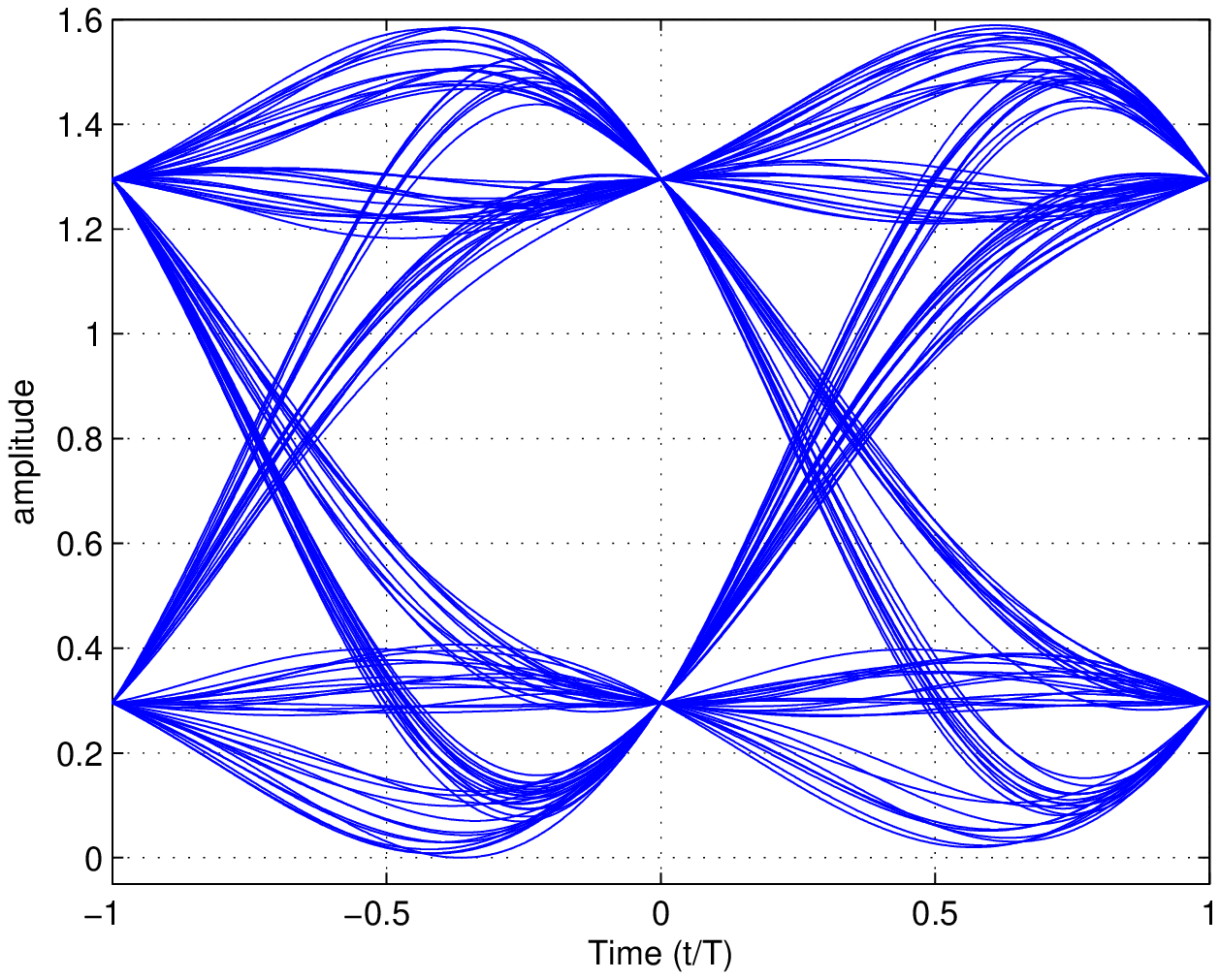}}Ê\\[-2ex]
\small (c) & \small (d)
\end{tabular}

\caption{Noise-free eye diagrams for (a) RC, (b) PL, (c) BTN, and (d) Xia pulses with OOK modulation ($\mathcal{C} = \{0,1\}$) and sampling receiver. All pulses have $\alpha=0.60$ and are normalized to have the same optical power $\bar{q}=1$.}
\label{eye-diagrams}
\end{center}
\end{figure}

\begin{figure}
%

%
\psfrag{10}{\footnotesize }
\psfrag{0.3}{\hspace{-4mm} 2}
\psfrag{0.1}{\footnotesize }
\psfrag{-0.3}{\hspace{-4mm} 0.5 }
\psfrag{-0.1}{\hspace{13mm} 1 }
\psfrag{-0.5}{\footnotesize }
\psfrag{B}{\footnotesize \hspace{-4mm}$BT_{\mathrm{b}}$}
\psfrag{Upsilon}{\footnotesize $\Upsilon$ (dB)}
\includegraphics[width=1\textwidth]{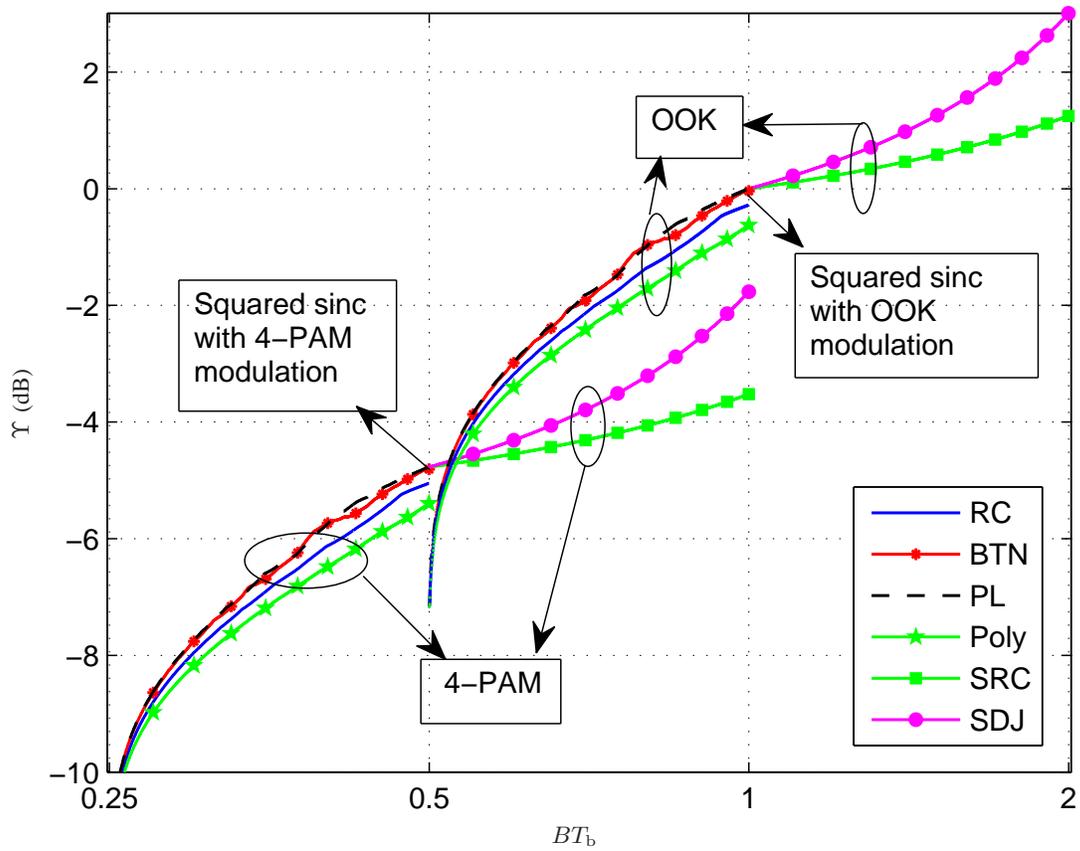}

\caption{\label{fig:OptPower_Eye}The  optical power gain $\Upsilon$ versus
normalized bandwidth $BT_{\mathrm{b}}$ for various Nyquist pulses with a sampling receiver. The  noiseless eye opening for all pulses
is equal. The curves for $BT_{\mathrm{b}}\geq 1$ agree with \cite{4132995}.}
%
\end{figure}

There is a compromise between bandwidth
and optical power gain,  due to the fact that  $\mu$ will be reduced by increasing the roll-off
factor (see Fig.~\ref{fig:RequiredDC}), whereas the required
bandwidth increases. The highest optical power gain for all  pulses will
be achieved when the roll-off factor $\alpha$ is one. The reason is that by increasing the roll-off factor, the required bias, which is the only  parameter in (\ref{eq:OptGainEye}) that depends on $\alpha$, decreases. The BTN  and the PL pulses have approximately similar optical power gain, and the Poly and RC pulses have smaller gains, due to higher $\mu$, which is also visible in the eye diagrams of Fig.~\ref{eye-diagrams}.  
 
Comparing the binary and 4-PAM cases for the same $\alpha$ and $\Delta a$, we can see in Fig.~\ref{fig:OptPower_Eye} that by using higher-order modulation formats, the optical power gain for all pulses decreases, since in (\ref{eq:OptGainEye}), $\mathbb{E}\left\{a_{k}\right\}$ and $\mu$ will increase. For $0.5<BT_{\mathrm{b}}<1$, the optical power gain for the best 4-PAM system with nonnegative Nyquist pulses is up to 2.39~dB less than the gain of the best OOK system with regular Nyquist pulses.


For any root-Nyquist pulse with a matched filter receiver, the minimum eye opening after filtering is given by (\ref{eq:Rec_MF}) as \vspace{-3mm}
\begin{equation}
\min_{a,a'\in \mathcal{C}, a\neq a'} \left|A \zeta\left(\mu Q(0)+a\Eq\right)-A\zeta\left(\mu Q(0)+a'\Eq\right)\right|=A\zeta\Delta a \Eq.\label{eq:MinEye_MFR}\end{equation}
Since the eye openings in \eqref{eq:MinEye_Sample} and \eqref{eq:MinEye_MFR} depend on the receiver filter gains $G(0)$ or $\zeta$, pulses should be compared using the same receiver filter. In particular, it is not relevant to compare the sampling receiver with matched filters in this context, since the outcome would depend on the ratio $G(0)/\zeta$, which can be chosen arbitrarily. This is the reason why  root-Nyquist pulses are not included in Fig.~\ref{fig:OptPower_Eye}.

\begin{figure}
\psfrag{10}{\footnotesize }
\psfrag{0.3}{\hspace{-4mm} 2}
\psfrag{0.1}{\footnotesize }
\psfrag{-0.3}{\hspace{-4mm} 0.5 }
\psfrag{-0.1}{\hspace{13mm} 1 }
\psfrag{-0.5}{\footnotesize }
\psfrag{B}{\footnotesize \hspace{-4mm}$BT_{\mathrm{b}}$}
\psfrag{Upsilon}{\footnotesize $\Upsilon$ (dB)}
\includegraphics[width=1\textwidth]{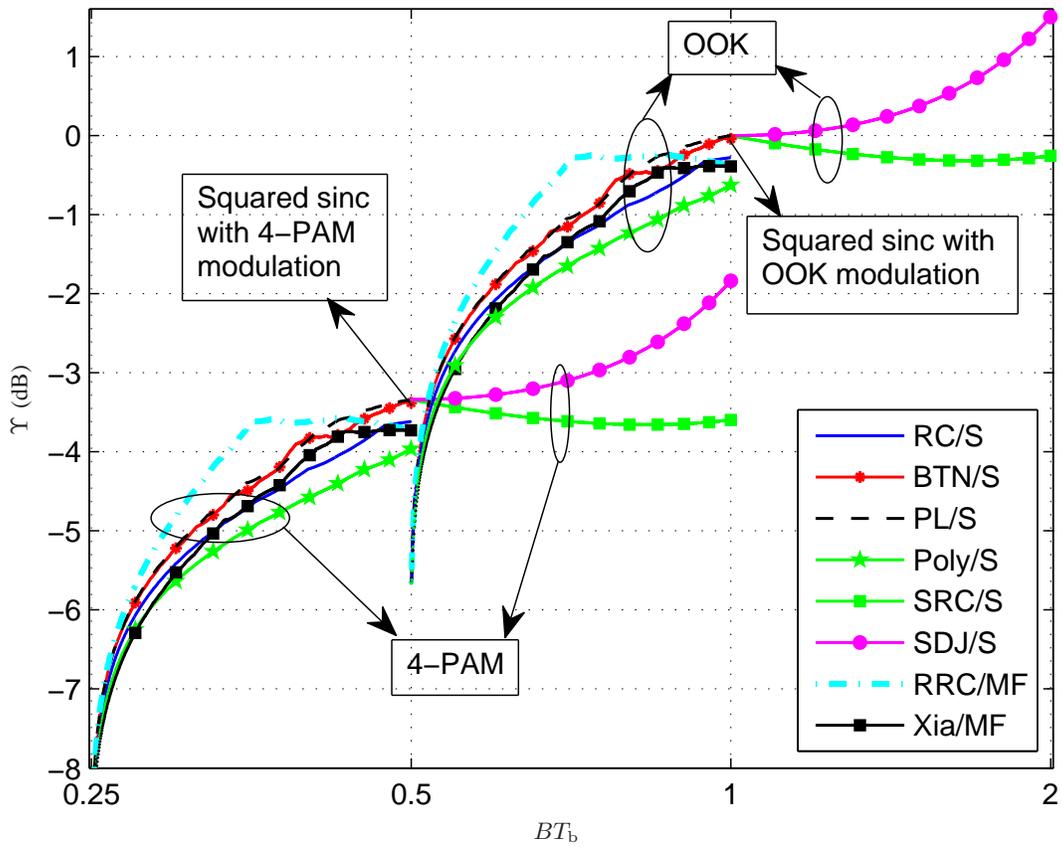}

\caption{\label{fig:OptPower_BER}The  optical power gain versus normalized bandwidth $BT_{\mathrm{b}}$ for various pulses with a sampling receiver (S) or matched filter receiver (MF). The SER for all pulses is $10^{-6}$.}
%
\end{figure}

It appears from Fig.~\ref{fig:OptPower_Eye} that the studied pulses become  more power-efficient when the bandwidth is increased. 
A higher bandwidth, however, for sampling receiver means that the receiver filter admits more noise, which reduces the receiver performance. In Fig.~\ref{fig:OptPower_BER}, we therefore  compare the average optical power
gain of Nyquist and root-Nyquist pulses, when the power is adjusted to yield a constant SER equal to $10^{-6}$. Since the amount of noise after the matched filter receiver does not depend on the bandwidth, we considered this fact as a potential advantage, and therefore included root-Nyquist pulses in the following analysis. Similarly to the previous case, the S2 pulse with OOK and sampling receiver is used as a baseline for comparison.

So far the analysis holds for a general $\mathcal{C}$. To find the optical power gain as a function of SER for the sampling receiver, we first apply a maximum likelihood detector to (\ref{eq:RecSamplesEQ}), assuming a special case in which  $\mathcal{C}$ is an $M$-PAM constellation, which yields the SER \cite [Sec.~9.3] {proakis2001digital}
\[
P_{\mathrm{err}}=2\frac{M-1}{M}Q\left(\frac{AG(0)\Delta a q(0)}{2\sqrt{G(0)^2N_{0}B}}\right)\]where
\[Q(x)=\frac{1}{\sqrt{2\pi}}\intop_{x}^{\infty}\exp\left(\frac{-x^{2}}{2}\right)dx\] is the Gaussian Q-function. 
 As a result,
\[
A=\frac{2}{\Delta a q(0)}Q^{-1}\left(P_{\mathrm{err}}\frac{M}{2\left(M-1\right)}\right)\sqrt{N_{0}B}\]
and
\begin{equation}
\frac{A_{\mathrm{ref}}}{A}=\frac{\Delta a q(0)}{\Delta a_{\mathrm{ref}}}\frac{Q^{-1}\left(P_{\mathrm{err}}\right)}{Q^{-1}\left(P_{\mathrm{err}}\frac{M}{2\left(M-1\right)}\right)}\sqrt{\frac{B_{\mathrm{ref}}}{B}},\label{eq:Ratio_Sampl}\end{equation}
where $B_{\mathrm{ref}}=1/T_\mathrm{b}$ is the bandwidth of the reference pulse. The optical power gain now follows from (\ref{eq:OptGainGeneral}).

For the matched filter receiver, by applying the maximum likelihood detector to (\ref{eq:Rec_MF}), the SER will be \cite [Sec.~9.3] {proakis2001digital}
\vspace{-4mm}
\begin{align}
P_{\mathrm{err}}=2\frac{M-1}{M}Q\left(\frac{A\Delta a\Eq\zeta}{2\sqrt{\frac{\zeta^2N_{0}\Eq}{2}}}\right)\nonumber\\
\qquad=2\frac{M-1}{M}Q\left(A\Delta a\sqrt{\frac{\Eq}{2N_{0}}}\right).\nonumber\end{align}
 As a result,
\[
A=\frac{1}{\Delta a}Q^{-1}\left(P_{\mathrm{err}}\frac{M}{2\left(M-1\right)}\right)\sqrt{\frac{2N_{0}}{\Eq}}\]and
\begin{equation}
\frac{A_{\mathrm{ref}}}{A}=\frac{\Delta a}{\Delta a_{\mathrm{ref}}}\frac{\sqrt{2}Q^{-1}\left(P_{\mathrm{err}}\right)}{Q^{-1}\left(P_{\mathrm{err}}\frac{M}{2\left(M-1\right)}\right)}\sqrt{\Eq B_{\mathrm{ref}}}.\label{eq:Ratio_MFRx}\end{equation}


In contrast to the case with equal eye openings (see Fig.~\ref{fig:OptPower_Eye}), Nyquist and root-Nyquist pulses can be compared with each other when the SER is kept constant, since neither (\ref{eq:Ratio_Sampl}) nor (\ref{eq:Ratio_MFRx}) depend on the filter gains $G(0)$ and $\zeta$.

By increasing the bandwidth, the gain for SRC decreases slightly, whereas it increases for SDJ, where $\mu=0$ for both cases. The reason is that for these pulses by increasing $\alpha$, both $\overline{q}$ and the ratio $A_{\mathrm{ref}}/A$ decreases.
 We observe that for the regular Nyquist pulses in Sec.~\ref{sec:Pulse-shaping}, the gain increases by increasing the bandwidth. The reason is that by increasing the roll-off factor, the required bias  decreases much faster (see Fig.~\ref{fig:RequiredDC}) than the speed of increase in bandwidth. The BTN and the PL pulses have approximately similar gain, and the gain of the RC and Poly pulses are always smaller than the gain of the other two pulses.

In case of  the matched filter receiver, the noise variance does not depend on bandwidth. As  a result, the ratio $A_{\mathrm{ref}}/A$ in (\ref{eq:Ratio_MFRx}) is not a function of the roll-off factor and the optical power gain only depends on the roll-off factor through its dependence on the required DC bias. In Fig.~\ref{fig:OptPower_BER}, the optical power gain of the RRC pulse increases for $0.5<BT_{\mathrm{b}}\le 0.71$, and a wide gap is maintained with respect to the Nyquist pulses. For $0.71<BT_{\mathrm{b}}\le 1$, since the required DC is slightly fluctuating, the same happens for the optical power gain of RRC, and the maximum optical power gain happens at $BT_{\mathrm{b}}=0.86$, where it is $\Upsilon = -0.22$~dB.
 The Xia pulse has a similar behavior, though it is not better than all Nyquist pulses.

 For $\alpha\rightarrow1$, the optical power gain  of the Xia, RC, and RRC pulses are approximately equal. In this case, although the output of matched filter will be equal to an RC pulse by either using RRC or Xia pulse, the performance will be different for other values of $\alpha$.

By increasing the modulation level from binary to 4-PAM, for the same $\alpha$ and $\Delta a$, the optical power gain for all pulses decrease, since the required DC bias and symbol average increase while the ratio $A_{\mathrm{ref}}/A$ decreases. For $0.5<BT_{\mathrm{b}} < 1$, the optical power gain of the regular Nyquist pulses and root-Nyquist pulses with OOK modulation is significantly more than the gain for the all nonnegative Nyquist pulses with 4-PAM.


When the
roll-off factor is equal to zero (i.e., the normalized bandwidth $BT_{\mathrm{b}}$ for the biased pulses with binary modulation is equal to 0.5 and for the biased pulses with 4-PAM is equal to 0.25), the regular Nyquist pulses discussed in Sec.~\ref{sec:Pulse-shaping} and the root-Nyquist pulses in Sec.~\ref{sec:RootPulse-shaping} will become equal to a sinc pulse with bandwidth $1/(2\Ts)$. As discussed in Sec.~\ref{sec:Required-DC-bias}, the required DC will
be infinite for the sinc pulse. Hence, the gain $\Upsilon$ will asymptotically go to $-\infty$ when $\alpha\rightarrow0$.

\vspace{-6mm}
\section{Conclusions \label{sec:CONCLUSIONS}}

In this work, a pulse shaping method for strictly bandlimited IM/DD systems
is presented, in which
the transmitted electrical signal must be nonnegative.
The proposed approach adds a constant DC bias to the transmitted signal, which allows a wider selection of transmitted pulses without violating the
nonnegativity constraint. This allows us to use Nyquist or root-Nyquist pulses for
ISI-free transmission, with narrower bandwidth compared to previous works. 
It is possible to transmit with a bandwidth equal to that of
ISI-free transmission in conventional coherent channels.

To compare our proposed
transmission schemes  with previously designed schemes and to see the effect of increasing the modulation level, we evaluated analytically the average optical
power versus bandwidth in two different scenarios.
The optimization of modulation formats means a tradeoff between the two components of the optical power: the constellation power, which carries the data and is similar to the coherent case, and the bias power, which is constant. We prove the somewhat unexpected results that for narrowband transmission ($B\Ts \le 1 $), the two powers balance each other perfectly, so that OOK and BPSK have identical performance regardless of the pulse.

In the first scenario, the Nyquist pulses are compared when the noise-free eye opening is equal for all the pulses and modulation formats. Of the studied pulses, the SDJ pulse with OOK is the best known, as previously shown in \cite{4132995} over $B T_\mathrm{b}\ge 1$.  At $0.5<B T_\mathrm{b}< 1$, the PL and BTN pulses with binary modulation have the best performance, being up to $2.39$~dB better than SDJ with 4-PAM modulation. Similarly, the 4-PAM BTN and PL pulses have highest gain over $0.25<B T_\mathrm{b}< 0.5$.

In the second scenario, all pulses have equal SER. Of the studied pulses, the SDJ with OOK modulation and sampling receiver has the highest gain for $B T_\mathrm{b}\ge 1$.  At $0.869<B T_\mathrm{b}< 1$, the binary PL pulse has the best performance, whereas for   $0.5<B T_\mathrm{b}\leq 0.869$, the RRC pulse with matched filter receiver achieves the highest gain. The gain of RRC in this scenario is up to $0.74$~dB over the best Nyquist pulse and $2.80$~dB over the best known results with unbiased  PAM.
It seems possible that further improvements can be achieved by utilizing the most recently proposed Nyquist pulses \cite{1369626,beaulieu2002better,chandan05,alexandru11,assimonis11}, or their corresponding root-Nyquist pulses, and carefully optimizing their parameters.

Extensions to $M$-PAM systems with $M>4$ are straightforward, in order to gain even more spectral efficiency at the cost of reduced power efficiency. This might be important for designing power- and bandwidth-efficient short-haul optical
fiber links (e.g., fiber to the home and optical interconnects) \cite{4528765,5875697} and diffuse indoor wireless
optical links \cite{gfeller2005wireless,kahn2002wireless,hranilovic2005design}.


\vspace{-8mm}
\bibliographystyle{IEEEtran}

\bibliography{MyDatabase_EA}
\newpage

\begin{table*}
\renewcommand\arraystretch{1.3}  
\small 
\begin{center}
\caption{Definitions of the studied Nyquist and root-Nyquist pulses.}
\label{pulse-def}
\begin{tabular}{|c|l|}
\hline
Pulse & Definition $q(t)$ \\
\hline\hline
RC & $\begin{cases}
\frac{\pi}{4}\sinc\left(\frac{t}{\Ts}\right),& t = \pm \frac{\Ts}{2\alpha}, \\
\sinc\left(\frac{ t}{\Ts}\right) \frac{\cos\left(\frac{\pi\alpha t}{\Ts}\right)}{1-(\frac{2\alpha t}{\Ts})^{2}},& \text{otherwise}
\end{cases}$ \\
BTN & $\sinc\left(\frac{t}{\Ts}\right)\frac{\frac{2\pi\alpha t}{\Ts \ln2}\sin\left(\frac{\pi\alpha t}{\Ts}\right)+2\cos\left(\frac{\pi\alpha t}{\Ts}\right)-1}{\left(\frac{\pi\alpha t}{\Ts \ln2}\right)^2+1}$ \\
PL & $\sinc\left(\frac{t}{\Ts}\right)\sinc\left(\frac{\alpha t}{\Ts}\right)$ \\
Poly &$\begin{cases}
1,& t = 0, \\
3\sinc\left(\frac{ t}{\Ts}\right) \frac{\sinc\left(\frac{\alpha t}{2\Ts}\right)^2-\sinc\left(\frac{\alpha t}{\Ts}\right)}{\left(\frac{\pi\alpha t}{2\Ts}\right)^{2}},& \text{otherwise}
\end{cases}$ \\
S2 & $\sinc^{2}\left(\frac{ t}{\Ts}\right)$ \\
SRC & $q_{\mathrm{RC}}^2(t)$, where $q_\mathrm{RC}$ is the RC pulse defined above \\
SDJ & $\left[\left(\frac{1-\alpha}{2}\right)\sinc\left(\frac{(1-\alpha) t}{\Ts}\right)+\left(\frac{1+\alpha}{2}\right)\sinc\left(\frac{(1+\alpha) t}{\Ts}\right)\right]^{2}$ \\
RRC & $\begin{cases}
1-\alpha+\frac{4\alpha}{\pi}, & t=0,\\
\frac{\alpha}{\sqrt{2}}\left[(1+\frac{2}{\pi})\sin(\frac{\pi}{4\alpha})+(1-\frac{2}{\pi})\cos(\frac{\pi}{4\alpha})\right], & t=\pm\frac{\Ts}{4\alpha},\\
\frac{\sin\left(\frac{\pi(1-\alpha)t}{\Ts}\right)+\frac{4\alpha t}{\Ts}\cos\left(\frac{\pi(1+\alpha)t}{\Ts}\right)}{\frac{\pi t}{\Ts}\left(1-\left(\frac{4\alpha t}{\Ts}\right)^{2}\right)}, & \text{otherwise}\end{cases}$ \\
Xia & $\sinc\left(\frac{t}{\Ts}\right)\frac{\cos\left(\frac{\pi \alpha t}{\Ts}\right)}{\frac{2\alpha t}{\Ts}+1}$ \\
\hline
\end{tabular}
\end{center}
\end{table*}

\begin{table*}
\small 
\caption{Parameters of all considered pulses. The energy $\Eq$ is relevant for root-Nyquist pulses only.}\label{tab:All_Info_Nyq}
\begin{center}

\begin{tabular}{|c|cc|cccc|}
\hline
Pulse & Nyquist & root-Nyquist & $\overline{q}$ & $q(0)$ & $B \Ts$ & $\Eq/\Ts$ \\
\hline
\hline
RC        & $\checkmark$ &         & $1$        & $1$        & $(1+\alpha)/2$ & \\
BTN        & $\checkmark$ &         & $1$        & $1$        & $(1+\alpha)/2$ & \\
PL        & $\checkmark$ &         & $1$        & $1$        & $(1+\alpha)/2$ & \\
Poly        & $\checkmark$ &         & $1$        & $1$        & $(1+\alpha)/2$ & \\
S2        & $\checkmark$ &         & $1$        & $1$        & $1$ & \\
SRC        & $\checkmark$ &         & $1-\alpha/4$    & $1$        & $1+\alpha$ & \\
SDJ        & $\checkmark$ &         & $1-\alpha/2$    & $1$        & $1+\alpha$ & \\
RRC        &        & $\checkmark$    & $1$        & $1-\alpha+4\alpha/\pi$ & $(1+\alpha)/2$ & $1$ \\
Xia        & $\checkmark$    & $\checkmark$    & $1$        & $1$        & $(1+\alpha)/2$ & $1$ \\
 \hline
\end{tabular}
\end{center}
\end{table*}

\end{document}